\documentclass{aastex631}
\usepackage[utf8]{inputenc}
\usepackage{graphicx}	% Including figure files
\usepackage{amsmath}	% Advanced maths commands
\usepackage{amssymb}	% Extra maths symbols

\usepackage{comment}

\newcommand{\VEC}[1]{\vec {#1} }
\begin{document}

%----%----%----%----%----%----%----%----%----%----%----%
%%%%%%%%%%%%%%%%%%%_TITLE_PAGE_%%%%%%%%%%%%%%%%%%%
% 
\title{
Correct criterion of crustal failure driven by intense magnetic stress in neutron stars
}
\date{\today}
\author[0000-0002-4402-3568]{Yasufumi Kojima}
\affiliation{
Department of Physics, Graduate School of Advanced Science and Engineering,\\
Hiroshima University, Higashi-Hiroshima, Hiroshima 739-8526, Japan}
%%

%%%%%%%%%%%%%%%%%%%%%%%%%%%%%%%%%%%%%%%%%%%%%%%%%%%%
\begin{abstract}
Magnetar outbursts are powered by an intense magnetic field.
The phenomenon has recently drawn significant attention because of a connection to some fast radio bursts that has been reported.
Understanding magnetar outbursts may provide the key to mysterious transient events.  
The elastic deformation of the solid crust due to magnetic field evolution accumulates over a secular timescale. 
Eventually, the crust fractures or responds plastically beyond a particular threshold.
Determination of the critical limit is required to obtain the shear strain tensor in response to magnetic stress.
In some studies, the tensor was substituted with an approximate expression determined algebraically from the magnetic stress.
This study evaluated the validity of the approximation by comparing it with
the strain tensor obtained through appropriate calculations. The differential equations for the elastic deformation driven by the magnetic field were solved.
The results indicated that the approximation did not represent the correct strain tensor value, both in
magnitude and spatial profile.
Previous evolutionary calculations based on spurious criteria
are likely to overestimate the magnitude of the strain tensor, and 
crustal failure occurs on a shorter timescale.
Therefore, revisiting evolutionary calculations using the correct approach is necessary.
This study is essential for developing the dynamics of crustal fractures and the magnetic-field evolution in a magnetar.
\end{abstract}
%%%%%%%%%%%%%%%%%%%%%%%%%%%%%%%%%%%%%%%%%%%%%%%
%
%\keywords{Neutron stars (1108); Compact objects (288); Magnetars (992); High-energy astrophysics (739)}
%
%
%\maketitle

%(1)%%%%%%%%%%%%%%%%%%%%%%%%%%%%%%%%%%%%%%%%%%%
\section{Introduction}
%%%%%%%%%%%%%%%%%%%%%%%%%%%%%%%%%%%%%%%%%%%%%%%
Magnetars are a peculiar class of neutron stars 
characterized by their variability which is driven by intense 
magnetic fields ~\citep{1992ApJ...392L...9D}.
The X-ray luminosity is very bright in the range 
$10^{32}$--$10^{36}$erg/s,
 which exceeds the spin-down luminosities of most sources.
Frequent bursting activity is also observed in soft gamma repeaters (SGRs).
 Short bursts typically last for 0.1--1 s, and the peak luminosity is $10^{39}$--$10^{41}$erg/s.
More energetic outbursts also occur, including extreme events such as a giant flare, in which energy over 
$10^{44}$erg is released in a moment~\citep[][for a review]
{2015RPPh...78k6901T,2017ARA&A..55..261K,2021ASSL..461...97E}.
Bursting events with larger energies are inevitably rare.
The energy supply for persistent emissions and transient bursts and flares
 observed in magnetars can be explained by
 the magnetic energy $\sim 10^{45}(B/ 10^{14}{\rm{G}})^2$erg
 stored in the interior star or exterior magnetosphere 
 with a typical field strength $B$.
 Magnetars probably possess more magnetic energy in their reserve because the interior toroidal magnetic field 
 is $10^2$ times stronger than that of the surface dipole,
as observationally suggested from the spin precession~\citep
{2014PhRvL.112q1102M,2016PASJ...68S..12M,2021MNRAS.502.2266M,2021ApJ...923...63M}.
Strong magnetic fields cause significant stress in the crust, and
the strain increases over time.
When the crust can no longer support the stress, it fractures, leading to 
soft gamma and X-ray emissions in the magnetosphere~\citep{1995MNRAS.275..255T}.
The crust-fracturing model resembles an earthquake, although with a different triggering force. 
Magnetic reconnection in the magnetosphere has been proposed as another outburst model in close analogy with solar flares~\citep{2003MNRAS.346..540L}.
The external magnetic fields become twisted over long periods owing to the plastic motion of the crust.
When the magnetosphere loses equilibrium beyond a certain threshold, a catastrophic transition results in an outburst or flare.
SGR bursts, earthquakes, solar flares, and other transient events
may share some common statistical properties.
One notable relationship is the power-law dependence of the number of bursts on 
their energy~\citep[][for SGR bursts]{1996Natur.382..518C,1999ApJ...526L..93G,2000ApJ...532L.121G},
known as the Gutenberg--Richter law for earthquakes.
Such statistical arguments regarding SGRs have attracted considerable attention recently
 because of a class of fast radio bursts (FRBs) which has been associated with a magnetar, 
 SGR J1935+2154~\citep{2020Natur.587...59B,2020Natur.587...54C}.
However, such a repeating source might not be representative,
and FRBs are, therefore, still mysterious events~\citep[e.g.,][for reviews]{
2020Natur.587...45Z,2022A&ARv..30....2P}.
The similar statistical properties between SGRs and repeating FRBs have been extensively studied
to explore their connection.
The scale-invariant energy feature was analyzed using the Tsallis q-Gaussian 
distribution~\citep{2021ApJ...920..153W,2022MNRAS.510.1801S}.
The correlation function between a pair of FRBs over a short timescale was explored, revealing a close similarity with that of earthquakes and the short bursts from the magnetar, SGR J1935+2154, but different from solar 
flares~\citep[][]{2023MNRAS.526.2795T,2024MNRAS.530.1885T}.
However, for longer intervals beyond the one-second threshold,
a unified scaling law in the temporal occurrence
 revealed a functional similarity
in functional form to solar flares rather than earthquakes~\citep{2024MNRAS.531L..57D}.
Arrival time patterns of magnetar bursts 
and repeating FRBs were compared within the framework of
 chaos and randomness~\citep{2024SciBu..69.1020Z,2024MNRAS.528L.133Y}.
In the energy domain, SGR bursts are consistent with 
 earthquakes, solar flares, and FRBs, whereas 
SGR bursts are less random than other transient phenomena.
Despite significant efforts, sufficient understanding of the similarity or dissimilarity between magnetar bursts and other transient phenomena is lacking.
The power-law form frequently emerges in various phenomena and
is characterized by a self-organized criticality (SOC) feature~\citep{1987PhRvL..59..381B},
that models complexities in nature without a detailed microscopic structure.
This concept has been applied to various phenomena, including
earthquakes~\citep{2002PhRvL..88q8501B},
solar flares~\citep{2015ApJ...814...19A,2016SSRv..198...47A},
gamma-ray bursts~\citep{2015MNRAS.453.2982D},
and magnetar bursts~\citep{2023ApJ...947L..16L}.
Detailed simulation on the microscopic scale is impossible, making the SOC concept a useful approach.
However, a microscopic understanding of various events is critical to distinguish between them.
How are crust fractures in a magnetar driven by intense magnetic stress 
~\citep{1995MNRAS.275..255T,2012MNRAS.427.1574L,
2015MNRAS.449.2047L,2017ApJ...841...54T}?
The failure of solid materials is determined by the strain tensor $\sigma _{ij}$.
The stable range by widely used criteria is determined by the condition that the difference between the maximum and minimum eigenvalues of $\sigma _{ij}$
(in the Tresca criterion) or the magnitude $ (\sigma _{ij} \sigma ^{ij}/2)^{1/2}$ (in the von Mises criterion)
is smaller than the critical number.
A crust responds elastically when its deformation is within the limit, 
 beyond which it cracks or responds plastically~\citep{2003ApJ...595..342J,
 2012MNRAS.427.1574L,2014ApJ...794L..24B}.
For the strain caused by the magnetic stress,
$\sigma$ is estimated by
the shear modulus $\mu$ and the change in the magnetic field $\delta B$,
as $\mu \sigma \sim  B \delta B/(4\pi)$ of the order of magnitude
 ~\citep[e.g.,][]{
1995MNRAS.275..255T,2021ApJ...919...89Y}.
In further detailed models, all components of the tensor $\sigma ^{ij}$ are necessary.
An approximate relation
$\sigma ^{ij} = -(2\mu)^{-1}\delta T_{\rm{mag}} ^{ij}$
was used to determine the elastic limit 
in the numerical simulations of the Hall magnetic-field evolution ~\citep{
2011ApJ...727L..51P,2013MNRAS.434..123V,2020ApJ...902L..32D},
where $\delta T_{\rm{mag}} ^{ij}$ denotes the change in the magnetic stress tensor.
They assumed that the crustal fracture occurred when 
the approximate strain tensor $\sigma ^{ij} = -(2\mu)^{-1} \delta T_{\rm{mag}} ^{ij}$
satisfied the breakup criterion.
They counted such events during their time-evolution simulation, 
and proposed a model for the observed burst rate based on the data. 
Another algebraic expression slightly different from
$-(2\mu)^{-1} \delta T_{\rm{mag}} ^{ij}$ was used 
in the numerical simulation of the critical transition from elastic to
plastic states~\citep{2019MNRAS.486.4130L,2021MNRAS.506.3578G}, and
for modeling crustal fractures
\citep[e.g.,][]{2015MNRAS.449.2047L,2016ApJ...824L..21L,2019MNRAS.488.5887S}.
The explicit expression for the elastic stress tensor algebraically obtained from the magnetic tensor
will be discussed in Section 2.
%%

%%%
However, the strain tensor cannot be determined using an algebraic equation.
Instead, it should be determined by solving the differential equations~\citep[e.g.,][]
{1959thel.book.....L}.
The elastic deformation of the neutron star has 
been calculated by solving the differential equations
for the nonmagnetized case~\citep[e.g.,][]
{2000MNRAS.319..902U,2020MNRAS.491.1064G,2021MNRAS.500.5570G,2022MNRAS.514.1628K},
and for the magnetized case~\citep{
2021MNRAS.506.3936K,2022MNRAS.511..480K,
2022ApJ...938...91K,2023ApJ...946...75K,2023MNRAS.519.3776F}.
The calculation is generally time-consuming.
When the approximate criterion is good, the previous evolutionary models without the trouble are justified. However, when the criterion is unjustified, evolutionary models should be revisited.
In this study, the validity of the algebraic expression 
$\sigma^{ij} = -(2\mu)^{-1}\delta T_{\rm{mag}} ^{ij}$
is considered by solving the differential equations for the elastic deformation driven by the
magnetic field.
However, the justification for this has not yet been discussed.
Therefore, this study is essential for 
developing the dynamics of crustal fractures and the magnetic-field evolution in a magnetar.
The remainder of this paper is organized as follows. Equations and models used 
are discussed in Section 2. 
The shear strain was numerically calculated
from magnetic field models. Section 3 compares the numerical results for the elastic stress tensor with those of the magnetic
tensor, which was approximately used as an elastic tensor in the literature. 
Finally, the conclusions are presented in Section 4.
%%

%(2)%%%%%%%%%%%%%%%%%%%%%%%%%%%%%%%%%%%%%%%%%%%
  \section{Mathematical formulation} 
%%%%%%%%%%%%%%%%%%%%%%%%%%%%%%%%%%%%%%%%%%%%%%%
%2.1%%%%%%%%%%%%%%%%%%%%%%%%%%%%%%%%%%%%%%%%%%%
  \subsection{Crustal model}
%%%%%%%%%%%%%%%%%%%%%%%%%%%%%%%%%%%%%%%%%%%%%%%
A crust of neutron stars from the core--crust interface at $r_{c}$ to
the surface at $R$ is considered, and the thickness is assumed to be $\Delta r/R=(R-r_{c})/R=0.1$.
The extremely thin outer crust is ignored; therefore, 
the mass density ranges from 
$\rho (r_c) =\rho_{c}$ $=1.4\times 10^{14}$ g cm $^{-3}$ 
at the core--crust boundary $r_{c}$ 
to the neutron-drip density 
$\rho(R)=\rho_{1}=4\times 10^{11}$ g cm$^{-3}$.
The hydrostatic equilibrium under Newtonian gravity is determined as follows:
 \begin{equation}
   \frac{d p}{dr} = -\rho g,
\label{hydroequil.eqn} 
 \end{equation}
 where $g{\VEC{e}}_{r}= \VEC{\nabla} \Psi_{\rm{G}}$ is the gravitational acceleration,
  which may be approximated as uniform throughout the crust
 because the crustal mass is extremely smaller than the total mass
 and the thickness was sufficiently small.
The reasonably approximated spatial profiles for the density $\rho$ and 
 pressure $p$ are obtained by
 \begin{equation}
    \rho =\rho_{c}[1-a(r-r_c)]^2,    ~~
     p =p_{c}[1-a(r-r_c)]^3,
\end{equation}
where $a = (1-(\rho_{1}/\rho_{c})^{1/2})/\Delta r$.
The pressure at $r=r_c$, $p_c = 10^{33}$ dyn cm$^{-2}$, and decreases 
$p(R) =p_{c}(\rho_{1}/\rho_{c})^{3/2}\approx 1.5 \times 10^{-4} p_{c}$
 $\approx 1.5 \times 10^{29} $ dyn cm$^{-2}$.
The gravitational acceleration is given by
  $g = 3ap_c/\rho_c$  $\approx 1.7 \times 10^{14}$ cm s$^{-2}$.
The shear modulus $\mu$ may be
approximated as a linear function of $\rho$,
which is overall-fitted to the results of a detailed calculation reported in a previous study
\citep[see Figure~43 in ][]{2008LRR....11...10C}.
Thus, $\mu$ is given by a radial function, approximated
in terms of the crustal spatial density profile
 \begin{equation}
    {\mu} = \mu_{c}[1-a(r-r_c)]^2,
%  \label{DFshMD.eqn} 
 \end{equation}
where $\mu_{c}=10^{30}~{\rm erg~cm}^{-3}$ at the core--crust interface.
%%

%2.2%%%%%%%%%%%%%%%%%%%%%%%%%%%%%%%%%%%%%%%%%%%
  \subsection{Elastic deformation}
%%%%%%%%%%%%%%%%%%%%%%%%%%%%%%%%%%%%%%%%%%%%%%%
The following two equilibrium states are considered: 
One is an initial state balanced without elastic deformation, and the other
is balanced with elastic stress because of the magnetic field evolution on a secular timescale.
The difference in the force-balance equation corresponding to the two states results in 
\begin{equation}
\nabla_{j} ( T_{\rm{hyd}} ^{ij}  +T_{\rm{elas}} ^{ij} +T_{\rm{mag}} ^{ij})
-\delta \rho{\VEC \nabla}\Phi_{\rm G} =0,
 \label{Forcebalance.eqn}
\end{equation}
where $T^{ij}$ is the stress tensor difference for each force~\footnote{
%%%
$\delta T^{ij}$ may be a better expression; however, 
the notation $T^{ij}$ is used without any confusion because 
only small first-order changes are considered.
}, respectively, given by
\begin{align}
&T_{\rm{hyd}} ^{ij}= -\delta p g^{ij},
  \label{T_hyd.eqn}\\
&T_{\rm{elas}} ^{ij}
 = 2\mu\sigma^{ij},
 \label{T_els.eqn}\\
&T_{\rm{mag}} ^{ij} =
 \frac{1}{4\pi}\left(
  B^{i}B^{j}-\frac{1}{2}g^{ij}B^2
  \right).
 \label{T_mag.eqn}
\end{align}
In Equation (\ref{Forcebalance.eqn}),  
 the change in the gravitational potential is ignored: $\delta \Phi_{\rm G}=0$,
 that is, the Cowling approximation is used.
The deviations are assumed to be so small that linear perturbation was applied.
Perturbations in the mass density $\delta \rho$
 and the pressure $\delta p$ are expressed by the Lagrange displacement ${\vec{\xi}}$;
\begin{equation}
\delta \rho = -
{\vec{\nabla}}\cdot (\rho {\vec{\xi}}) ,
~~
 \delta p = -({\vec{\xi}} \cdot {\vec{\nabla}})p
-\Gamma p ({\vec{\nabla}}\cdot {\vec{\xi}} ),
\end{equation}
where an adiabatic change in pressure is assumed, and
$\Gamma$ denotes the adiabatic index.
The elastic stress tensor (\ref{T_els.eqn}) is given by
the shear modulus $\mu$ and trace-free strain tensor $\sigma_{ij}$;
\begin{equation}
\sigma_{ij} =\frac{1}{2}
\left( \nabla_{i} \xi_{j}+
\nabla_{j} \xi_{i}  \right)
-\frac{1}{3} g_{ij} \nabla_{k} \xi^{k} ,
\label{straintensor.eqn}
\end{equation}
where $g_{ij}$ denotes a three-dimensional metric.
Spherical coordinates ($r, \theta, \phi$) are used, and
the explicit components of a vector and
 tensor are expressed in an orthonormal basis 
$(\partial_{r}, r^{-1}\partial_{\theta} ,(r\sin \theta)^{-1} \partial_{\phi})$.
Axial symmetry,  $\partial _{\phi}=0$  is assumed for the magnetic field and the displacement to simplify the calculation.
The displacement vector ${\VEC{\xi}}$ is
expanded in terms of the Legendre polynomials $P_l(\cos\theta)$ and
radial functions $R_{l}(r)$, $U_{l}(r)$, and $V_{l}(r)$;
\begin{equation}
[\xi_{r}, \xi_{\theta}, \xi_{\phi} ]
=\sum_{l}
\left[  r R_{l} P_{l},~ r U_{l} P_{l, \theta},~
 - rV_{l}P_{l, \theta} \right],
\end{equation}
where radial functions 
 $U_{l}$ and $V_{l}$ are meaningful for the index $l \ge 1$; however, it is convenient to introduce $U_{0}=V_{0}=0$.
Tensor components $T_{ij}$ are generally decomposed 
in terms of the spherical tensor harmonics and radial functions
$a_{l}(r)$, $b_{l}(r)$, $c_{l}(r)$, $d_{l}(r)$, $f_{l}(r)$ and $g_{l}(r)$ as
\begin{align}
T_{ij} = \sum_{l}
\begin{pmatrix}
   a_{l}P_{l} &   b_{l} P_{l, \theta} &  -c_{l} P_{l, \theta}\\
  b_{l} P_{l, \theta} & f_{l}W_{l} +g_{l} P_{l} &
 d_{l}W_{l} \\
  -c_{l} P_{l, \theta}  & d_{l}W_{l}  &  -f_{l}W_{l} + g_{l}P_{l} 
\end{pmatrix},
\label{TijSum.eqn}
\end{align}
where $W_{l}(\theta)$ is an angular function defined as
\begin{equation}
W_{l} =\sin \theta \left( \frac{P_{l,\theta}}{\sin \theta} \right)_{,\theta}.
%
% \label{DFWWlm.eqn} 
\end{equation}
The lower limit of $l$ in the sum 
(\ref{TijSum.eqn}) is limited by the fact that
$P_{0,\theta}=0$ and $W_{0}=W_{1}=0$.
The expression with the index $l$ may be written by using 
$b_{0}=c_{0}$$=d_{0}=d_{1}=f_{0}=f_{1}=0$.
The magnetic field evolution between equilibria is not explicitly considered,
because the magnetic field necessary for elastic deformation is obtained only by numerical calculations.
Moreover, the change during the evolution depends on the initial conditions. 
The initial magnetic field used in previous numerical works was chosen arbitrarily because the force-balance 
including the magnetic force was ignored ~\citep{
2011ApJ...727L..51P,2013MNRAS.434..123V,2020ApJ...902L..32D,2021MNRAS.506.3578G}.
The magnitude of the magnetic field is so small that the effect is neglected in the force-balance equation, 
whereas the magnetic field is highly constrained by considering equilibrium condition.
By avoiding complicated calculations, the magnetic stress tensor $T_{\rm{mag}} ^{ij}$ in Equation~(\ref{T_mag.eqn}) 
is analytically given and the displacement ${\vec{\xi}}$ is calculated.
Unlike the forms of $T_{\rm{hyd}} ^{ij}$ and $T_{\rm{elas}} ^{ij}$,
$T_{\rm{mag}} ^{ij}$ is the quadratic of $B^i$.
This study does not explicitly split $B^i$ into its initial value and an assumed slight variation; instead, a small change in the stress tensor given by the form of $T_{\rm{mag}} ^{ij}$ is assumed.
The radial functions in Equation~(\ref{TijSum.eqn})
are obtained for the specified magnetic-field model.
A potential limitation of using the assumed magnetic field is that the magnetic stress tensor used in 
this calculation may significantly differ from that of evolved field. 
This occurrence is expected to be rare, but should be checked in future works.
%%

%2.3%%%%%%%%%%%%%%%%%%%%%%%%%%%%%%%%%%%%%%%%%%%
  \subsection{Axial perturbation}
%%%%%%%%%%%%%%%%%%%%%%%%%%%%%%%%%%%%%%%%%%%%%%%
The axial perturbation $(\xi_{r}=\xi_{\theta}=0, \xi_{\phi} \ne 0)$ is assumed, in which the density and pressure perturbations are decoupled.
After separating the angular part, the $\phi$ component of 
Equation~(\ref{Forcebalance.eqn}) is reduced to an ordinary differential equation~\citep[see e.g.,][]
{2021MNRAS.506.3936K,2022MNRAS.511..480K}
\begin{equation}
  \left(\mu r^4 V_{l} ^{\prime}  
  +   c_{l} r^3 \right)^{\prime}
  -(l(l+1)-2)r^2( \mu V_{l}  -d_{l})=0.
 \label{axV2.dfeqn}
 \end{equation}
 Here, prime ($^\prime$) denotes a derivative with respect to $r$. 
For $l=1$, 
Equation (\ref{axV2.dfeqn}) is analytically calculated as
$\mu r V_{1} ^{\prime} + c_{1}=0$, 
where the integration constant is fixed by the boundary condition.
The $r\phi$ and $r\phi$ components are relevant to the axial perturbation, and the elastic stress is explicitly written as 
\begin{equation}
T_{\rm{elas}} ^{r\phi}=2\mu \sigma ^{r\phi}=- \mu \sum_{l} r V_{l} ^{\prime}P_{l,\theta},
~~
T_{\rm{elas}} ^{\theta \phi}=2\mu \sigma ^{\theta \phi}= -\mu \sum_{l} V_{l} W_{l}.
\label{tmnaxelas.eqn}
\end{equation}
The corresponding components of the magnetic stress are given by Equation~(\ref{TijSum.eqn})
\begin{equation}
 T_{\rm{mag}} ^{r\phi}= -\sum_{l} c_{l} P_{l, \theta},
~~
 T_{\rm{mag}} ^{\theta \phi}= \sum_{l} d_{l} W_{l}.
 \end{equation}
The condition $T_{\rm{elas}} ^{i \phi} + T_{\rm{mag}} ^{i \phi} =0$ ($i=r, \theta$) leads to
 $\mu  r V_{l} ^{\prime} +c_{l} =0$ and $\mu V_{l}-d_{l} =0$.
The function $V_{l}$ cannot satisfy these equations identically because
 $V_{l}$ is determined using Equation~(\ref{axV2.dfeqn}) for the independent functions
  $c_{l}$ and $d_{l}$.
The relation 
$T_{\rm{elas}}^{i \phi} + T_{\rm{mag}} ^{i \phi} =0$
 ($i=r, \theta$) cannot be satisfied, and 
 the accuracy of the approximation is examined 
 by solving Equation~(\ref{axV2.dfeqn}) in Section 3.
%%

%2.4%%%%%%%%%%%%%%%%%%%%%%%%%%%%%%%%%%%%%%%%%%%
  \subsection{Polar perturbation}
%%%%%%%%%%%%%%%%%%%%%%%%%%%%%%%%%%%%%%%%%%%%%%%
For polar perturbations
($(\xi_{r},\xi_{\theta}) \ne 0, \xi_{\phi} =0$),
tensor components 
$T^{rr}, T^{\theta \theta}, T^{\phi \phi}$, and $ T^{r\theta}$
are relevant, and 
the density and pressure changes are generally associated with the perturbations.
In the Cowling approximation, the perturbation equations
were derived~\citep[see e.g.,][]{1988ApJ...325..725M,2000MNRAS.319..902U}.
The magnetic stress terms ($a_{l}$, $b_{l}$, $f_{l}$, and $g_{l}$) are included in previous
studies. However, the variables and notations used herein are
slightly different from those reported in previous studies.
The $r$ and $\theta$ components of Equation~(\ref{Forcebalance.eqn})
are reduced to the following set of fourth-order differential equations:
\begin{align}
 R_{l} ^{\prime} &= \frac{l(l+1) \alpha_{2}}{\alpha_{3} r} U_{l}
 -\left( \frac{3\Gamma}{\alpha_{3}} + 
 \frac{r p^{\prime}}{\alpha_{3}p} \right)\frac{R_{l}}{r}+
 \frac{1}{\alpha_{3}pr^4} (X_{l} - a_{l}r^3),
  \label{plRUXY1.dfeqn}
  \\
U_{l} ^{\prime} &= -\frac{R_{l}}{r} +
 \frac{1}{\mu r^4} (Y_{l} - b_{l}r^3),
  \label{plRUXY2.dfeqn}
  \\
X_{l} ^{\prime} &= \frac{l(l+1)}{r} Y_{l}
+\left( \frac{3\Gamma}{\alpha_{3}} + 
 \frac{r p^{\prime}}{\alpha_{3}p} \right) 
 \left(
 \frac{X_{l}}{r}
    - 2l(l + 1) \mu r^2 U_{l} -a_{l}r^2 \right)
    \nonumber\\
  &  - \left[\rho^{\prime} g r^2 +
   \frac{(rp^{\prime})^2}{\alpha_{3}p}
  - 4 \mu \left(
  \frac{3\Gamma}{\alpha_{3}} + 
 \frac{2 r p^{\prime}}{\alpha_{3}p} 
  \right)
  \right]r^2 R_{l} + r^2( a_{l}+2g_{l} ),
     \label{plRUXY.dfe3qn}
    \\
Y_{l} ^{\prime} &= -
\frac{ \alpha_{2}}{\alpha_{3} r} X_{l} +
(l(l + 1) - 2)\left( \mu r^2 U_{l} +f_{l} r^2
\right)
    \nonumber\\
    &
    -\left( \frac{6\Gamma}{\alpha_{3}} + 
 \frac{2r p^{\prime}}{\alpha_{3}p} \right)\mu r^2R_{l} 
 + \frac{3l(l+1)\Gamma}{\alpha_{3}} \mu r^2 U_{l} 
     + \frac{r^2}{2}
     \left[
     \left(\frac{3\Gamma}{\alpha_{3}}-1\right)a_{l}-2g_{l}\right].
      \label{plRUXY4.dfeqn}
%%%%%%%%%%%%%%%%%%%%%%%%%%%%%%%%%%%%%%%%%%
 \end{align}
where
\begin{equation}
    \alpha_{2}= \Gamma - \frac{2\mu}{3p}
    \approx \Gamma,
    ~~~
    \alpha_{3}= \Gamma + \frac{4\mu}{3p}
    \approx \Gamma .
\end{equation}
These functions are approximated as $\Gamma$ because $\mu \ll p$.
The adiabatic constant is assumed to be constant and
$\Gamma = 1.5$ is used in the numerical calculations described in Section 3.
The $r\theta$ component of the elastic stress is expressed as
\begin{equation}
 T_{\rm{elas}} ^{r \theta} =2 \mu \sigma ^{r \theta} = \sum_{l} 
\mu (r U_{l} ^{\prime} +  R_{l})  P_{l, \theta}
= \sum_{l} \left(\frac{Y_{l}}{r^3}-b_{l} \right) P_{l, \theta}
= -T_{\rm{mag}} ^{r\theta}
+\sum_{l} \left(\frac{Y_{l}}{r^3} \right) P_{l, \theta},
\label{tmn12elas.eqn}
\end{equation}
where Equations (\ref{TijSum.eqn}) and (\ref{plRUXY2.dfeqn}) are used.
The relation
$T_{\rm{elas}} ^{r \theta} +T_{\rm{mag}} ^{r\theta}=0$ holds,
when $ Y_{l}=0$.
This condition is generally not satisfied.
The relationship between the shear and magnetic tensors in the diagonal part is complicated.
The trace of elastic stress tensor $T_{\rm{elas}}^{ij} $ is zero, whereas 
that of $T_{\rm{mag}} ^{ij}$ is nonzero.
Therefore, the relation
$T_{\rm{elas}} ^{ij}  =-T_{\rm{mag}} ^{ij}$
is assumed only in the off-diagonal region \citep{
2011ApJ...727L..51P,2013MNRAS.434..123V,2020ApJ...902L..32D}.
In order to remove the discrepancy in the trace, 
the traceless form ${\hat{T}}_{\rm{mag}} ^{ij}$ for $B^i$ is proposed
to obtain an approximate elastic stress tensor\citep{
2015MNRAS.449.2047L,2016ApJ...824L..21L,2019MNRAS.486.4130L,
2019MNRAS.488.5887S,2021MNRAS.506.3578G};
$T_{\rm{elas}} ^{ij} =-{\hat{T}}_{\rm{mag}} ^{ij}$
where
\begin{equation}
{\hat{T}}_{\rm{mag}} ^{ij}  =
 \frac{1}{4\pi}\left(
  B^{i}B^{j}-\frac{1}{3}g^{ij}B^2
  \right).
%\label{T_mag.eqn}
\end{equation}
Moreover, pressure perturbation also affects the relation
$T_{\rm{elas}} ^{ij} =-T_{\rm{mag}} ^{ij}$ or
$T_{\rm{elas}} ^{ij} =-{\hat{T}}_{\rm{mag}} ^{ij}$.
Therefore, the relation does not hold in this exact form, and 
the accuracy of the approximation may be verified
by using the numerical calculations described in Section 3.
However, this comparison may not be appropriate,
e.g., for $T^{rr}$, which depends on the trace of $T^{ij}$.
Therefore, two trace-free combinations,
$T^{\theta\theta} -T^{\phi \phi}$ and
$2T^{rr}-T^{\theta\theta} -T^{\phi \phi}$ are used to compare the diagonal part.
They are unchanged if $T_{\rm{mag}} ^{ij}$ is replaced by
${\hat{T}}_{\rm{mag}} ^{ij}$.
%%

%2.5%%%%%%%%%%%%%%%%%%%%%%%%%%%%%%%%%%%%%%%%%%%
\subsection{Boundary conditions}
%%%%%%%%%%%%%%%%%%%%%%%%%%%%%%%%%%%%%%%%%%%%%%%
The radial component of the total stress tensor 
$\sum T^{ri}~(i=r, \theta, \phi)$, i.e., 
the traction must be continuous across the surfaces at $r=r_c$ and 
$R$~\citep[see e.g.,][]{1988ApJ...325..725M,2000MNRAS.319..902U}.
For $i=\phi$, the total $T_{\rm{elas}} ^{r\phi} +T_{\rm{mag}} ^{r\phi}$
must be continuous. 
The magnetic component is assumed to be continuous on both sides.
This case corresponds to the axial perturbation with $\xi_{\phi}$, and
the boundary conditions for the radial function $V_{l}$ at $r= r_c$ and $R$ can be written as follows:
\begin{equation}
 V_{l}^{\prime}=0.
     \label{boundaryAX.eqn}
\end{equation}

Pressure perturbation is involved in polar perturbation. 
Both $T_{\rm{hyd}}^{ri}$ and $ T_{\rm{mag}}^{ri}$
 are assumed to be continuous to
the interior core $(r\le r_c)$ and the exterior $(r\ge R)$.
The continuity of the $rr$ and $r\theta$ components of the traction at
$r= r_c$ and $R$ leads to
\begin{align}
& 2r R_{l} ^{\prime} + l(l+1) U_{l}=0,
  ~\Longleftrightarrow~   X_{l} -(rp^{\prime} +3\Gamma p)r^3R_{l} 
  +\frac{3}{2}l(l+1)\Gamma p r^3 U_{l}=ar^3,
%
%\label{boundaryPO1.eqn}
\nonumber\\
  &r U_{l} ^{\prime} + R_{l}=0,
  ~~~~~~~~~~~~\Longleftrightarrow~ Y_{l}=br^3,
\label{boundaryPO2.eqn}
\end{align}
where Equations~(\ref{plRUXY1.dfeqn}) and (\ref{plRUXY2.dfeqn}) 
are used to obtain the second expression.
%

%2.6%%%%%%%%%%%%%%%%%%%%%%%%%%%%%%%%%%%%%%%%%%%
  \subsection{Magnetic field in the crust}
%%%%%%%%%%%%%%%%%%%%%%%%%%%%%%%%%%%%%%%%%%%%%%%
%%%%%%%%%%%%%%%%%[FIG0]%%%%%%%%%%%%%%%%%%%%%%%%
\begin{figure}\begin{center}
\includegraphics[width=0.8\columnwidth]{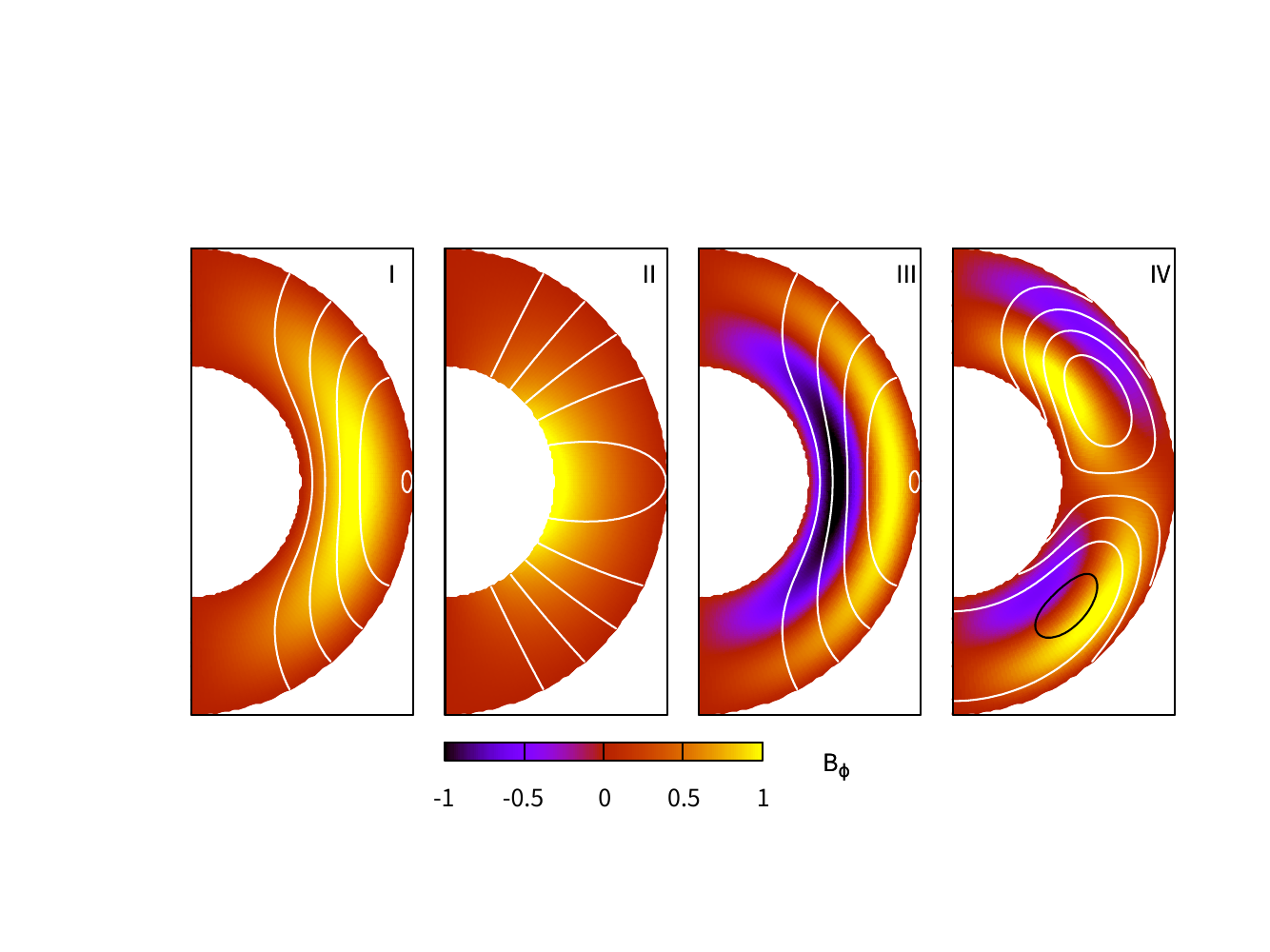}
\caption{ 
\label{Fig.Bgeometry}
Magnetic field geometry in the crust for four models.
Magnetic function $\Psi(r,\theta)$ 
for a poloidal field is represented by contours,
and the field strength of the toroidal component $B_{\phi}$, 
normalized by the maximum of $|B_{\phi}|$, is represented by colors.
Crustal region is five times enlarged for display.
}
\end{center}\end{figure}
%%%%%%%%%%%%%%%%%%%%%%%%%%%%%%%%%%%%%%%%%%%%%%%

%%
It is assumed that the external field is 
a dipole in vacuum $(r\ge R)$.
The poloidal field is described as
${\vec{B}}_{p}={\vec{\nabla }} \times (\Psi  {\vec{e}}_{\phi}/(r\sin \theta))$
in terms of the magnetic function $\Psi$.
The explicit form of $\Psi$ is given by
\begin{equation}
\Psi = -\frac{B_{0}R^3}{2r}
\sin \theta P_{1,\theta} = \frac{B_{0}R^3}{2r} \sin^2 \theta, 
\end{equation}
where $B_{0}$ is the field strength at the magnetic pole
on the surface $r=R$ and is hereafter used as the normalization constant. 
The toroidal component $B_{\phi}$ disappears on the exterior($r\ge R$).
The following four models are considered for the crustal magnetic field, 
which is connected to the dipole in a vacuum at $r=R$. 
Model I describes the dipolar magnetic field that is expelled from the core.
The magnetic function and toroidal component are given by
\begin{equation}
    \Psi ^{\rm{I}} = -\frac{B_{0}}{2} \left[ 2R^2 - rR 
-\left( R-r \right)^2
\frac{R(R+ \Delta r)}{\Delta r ^2}
\right]  \sin \theta P_{1,\theta},
    ~~
    B^{\rm{I}} _{\phi} = \frac{B_{0} R}{r}
    \sin \left( \frac{\pi(r-R)}{\Delta r}\right)
      P_{1,\theta} .
\label{dipolBI.eqn}
\end{equation}
These functions are displayed in the left panel of Figure~\ref{Fig.Bgeometry}.
Owing to the steep change in $\Psi$ in the radial direction, 
$|B_{\theta}|$ increases toward the inner boundary $r_c$
approximately ten times more than the
other components, $|B_{r}|$ and $|B_{\phi}|$. 
%

%%%
In Model II, the poloidal and toroidal components of the 
 magnetic field are smoothly connected 
 from the core to the crust, as displayed in
 the second panel of Figure~\ref{Fig.Bgeometry}.
The field is described as follows:
\begin{equation}
 \Psi ^{\rm{II}} =
  -\frac{B_{0}}{2} \left[ 2R^2 - rR 
+\left( R-r \right)^2\right]
\sin \theta P_{1,\theta},
    ~~
    B^{\rm{II}} _{\phi} = 
  \frac{B_{0} R}{r} 
    \left( \frac{r-R}{\Delta r}\right)
     P_{1,\theta} .
     \label{dipolBII.eqn}
\end{equation}
All components of the magnetic field
are comparable in the magnitude $\sim B_{0}$,
although they depend on the spatial position.
In Model III, the poloidal
magnetic field is the same as that of
Model I, but the toroidal component
increases by one order of magnitude.
Furthermore, a node exists in the radial
direction, that is,
$B_{\phi}$ changes in the direction, as shown 
in the third panel of Figure~\ref{Fig.Bgeometry}.
The magnetic field is given by
\begin{equation}
    \Psi ^{\rm{III}} = \Psi ^{\rm{I}}, 
    ~~
    B^{\rm{III}} _{\phi} =
    10 \times  \frac{B_{0} R}{r}
    \sin \left( \frac{2 \pi(r-R)}{\Delta r}\right)
 P_{1,\theta} .
 \label{dipolBIII.eqn}
\end{equation}
 The magnitude of the magnetic field 
is given by $|B_{\theta}| \sim |B_{\phi}| \sim 10 \times |B_{r}|$.

%%%
Finally, Model IV is a combination of $l=1$ and $l=2$.
A confined component with an angular dependence of $l=2$
is added to that of Model I, as follows:
\begin{equation}
   \Psi ^{\rm{IV}} = \Psi ^{\rm{I}} +
  \frac{B_{0}R^2}{2} \sin\left( \frac{r-R}{\Delta r}\right)
\sin \theta P_{2,\theta},
    ~~
    B^{\rm{IV}} _{\phi} = B^{\rm{I}} _{\phi} 
- \frac{B_{0} R}{r}
    \sin \left( \frac{2\pi(r-R)}{\Delta r}\right)
      P_{2,\theta} .
      \label{dipolBIV.eqn}
\end{equation}
The toroidal magnetic field with $l=2$ yields 
a node in the radial direction.
These functions (Equations
(\ref{dipolBI.eqn})--(\ref{dipolBIV.eqn})) are examples for convenience to provide the magnetic stress 
$T_{\rm{mag}}^{ij}$ which drives the elastic deformation.
As discussed in Section 2.2,
$T_{\rm{mag}}^{ij}$ is expressed by $B^{i}$, and
the models are chosen from various mathematical possibilities with a low multipole index $l$.
The astrophysical situation leading to the  magnetic stress $T_{\rm{mag}}^{ij}$ however is not considered.
These models are summarized in Table~\ref{table1:mylabel}
as shown in Figure~\ref{Fig.Bgeometry}.
The maximum value of each component and the average field strength are also added, calculated in the crust as
$B_{\rm{av}}^2\equiv$$\int B^2 d^3x/\int d^3x$.
%%

%%%%%%%%%%%%%%%%%%%[TABLE1]%%%%%%%%%%%%%%%%%%%%%
\begin{table}
\centering\begin{tabular}{llccccc}
Model&  Magnetic field geometry&   
 Equation&  $|B_{r}/B_0|_{\rm{max}}$&
 $|B_{\theta}/B_0|_{\rm{max}}$& 
 $|B_{\phi}/B_0|_{\rm{max}}$ &  $(B_{\rm{av}}/B_{0})^2 $\\
\hline \hline
%%%%
 I & Dipole expelled from core&
 (\ref{dipolBI.eqn})&
1.02& 11.7& 1.05 & 26.5\\
%%%%%%%
  II & Dipole penetrated to core&
     (\ref{dipolBII.eqn})&
     1.37& 0.67 & 1.11& 0.93\\
%%%%%%%
III & Dipole with strong toroidal field with a node&
   (\ref{dipolBIII.eqn})&
   1.02& 11.7& 10.3& 63.0\\
%%%%%%%
IV & Mixed toroidal-poloidal field with $l=1$ and 2&
    (\ref{dipolBIV.eqn})&
    4.18& 34.5& 2.17 & 192\\
%%%%%%%%%%%%%%%%%%%%%%%%%%%%%%%%%%%%%%%%%%%%%%%
    \hline \hline
    \end{tabular}
    \caption{Description of magnetic model}
\label{table1:mylabel}
\end{table}
%%%%%%%%%%%%%%%%%%%%%%%%%%%%%%%%%%%%%%%%%%%%%%%

%(3)%%%%%%%%%%%%%%%%%%%%%%%%%%%%%%%%%%%%%%%%%%%
   \section{Results}
%%%%%%%%%%%%%%%%%%%%%%%%%%%%%%%%%%%%%%%%%%%%%%%
%3.1%%%%%%%%%%%%%%%%%%%%%%%%%%%%%%%%%%%%%%%%%%%
  \subsection{Comparison between elastic and magnetic tensors}
%%%%%%%%%%%%%%%%%%%%%%%%%%%%%%%%%%%%%%%%%%%%%%%
%%%%%%%%%%%%%%%%%[FIG2]%%%%%%%%%%%%%%%%%%%%%%%%
\begin{figure}\begin{center}
\includegraphics[width=0.7\columnwidth]{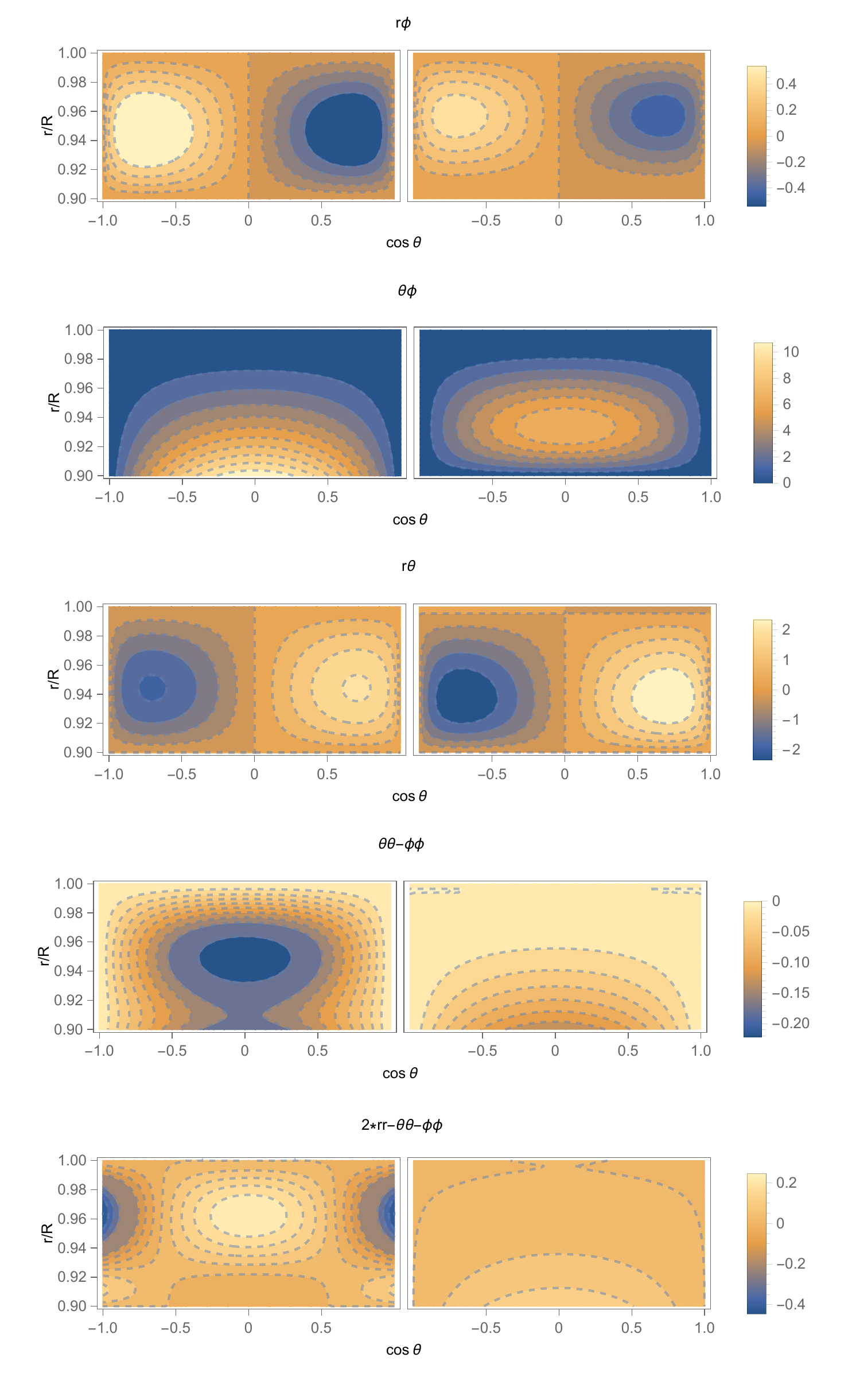}
\caption{ 
 \label{Fig.result1x5}
Display of tensor components via contours in the $\cos \theta$-$r/R$ plane for Model I.
The elastic stress $T_{\rm{elas}} ^{ij}$ (left panel)
is compared with the magnetic one $ -T_{\rm{mag}} ^{ij}$ (right panel).
From the top to bottom columns,
$T ^{r\phi}, T^{\theta \phi}, T^{r\theta}$,
$T^{\theta \theta}-T^{\phi \phi}$, and  
$2T^{rr}-T^{\theta \theta}-T^{\phi \phi}$ components are shown.
In the fourth and fifth panels, the magnetic part is so large that 
their maximum is adjusted by multiplying by a factor of $10^{-3}$. 
}
\end{center}\end{figure}
%%%%%%%%%%%%%%%%%%%%%%%%%%%%%%%%%%%%%%%%%%%%%%%

The displacement vector ${\vec{\xi}}$ is obtained by numerically 
solving Equations~(\ref{axV2.dfeqn}) and 
(\ref{plRUXY1.dfeqn})--(\ref{plRUXY4.dfeqn})
for the source terms $T_{\rm{mag}} ^{ij}$, given by
the model described in Section 2.6.
The strain tensor~(\ref{straintensor.eqn}) is calculated,
and the resultant elastic stress $T_{\rm{elas}} ^{ij} =2\mu \sigma^{ij}$ 
is compared with $-T_{\rm{mag}} ^{ij}$ in Figure~\ref{Fig.result1x5}.
The latter is proposed as an approximate form 
for the elastic stress tensor.
The magnitudes of the tensor components are indicated by the contours 
in the $\cos \theta$--$r/R$ plane
for the magnetic-field Model I.
Different components are normalized by a common factor such that
the relative comparisons between them are meaningful.
The result suggests that
the off-diagonal components of the elastic stress tensor may be approximated as
 $T_{\rm{elas}} ^{ij} \sim -T_{\rm{mag}} ^{ij}$,
although their maximum positions differ
in the $\theta \phi$ component (second panel of
Figure~\ref{Fig.result1x5}).
The difference in the radial direction
originates from the following boundary values:
$T_{\rm{mag}} ^{\theta \phi} =0$; however, $T_{\rm{elas}}^{\theta \phi}\ne 0$ at $r=r_c$.
The boundary conditions for the differential equation
(\ref{axV2.dfeqn}) are $V_{2} ^{\prime} =0$
(Equation~(\ref{boundaryAX.eqn})),
and $T_{\rm{elas}} ^{\theta \phi}=-V_{2}W_2$
(Equation~(\ref{tmnaxelas.eqn})) is determined posteriorly and is generally nonzero at $r=r_c$.
As discussed in Section 2.4, 
$T_{\rm{elas}} ^{ij}$ does not agree with $-T_{\rm{mag}} ^{ij}$ in the diagonal components;
the trace of $T_{\rm{elas}} ^{ij}$ is zero, whereas
that of $T_{\rm{mag}} ^{ij}$ is nonzero.
Therefore, $T_{\rm{elas}} ^{ij}$ is compared with $-T_{\rm{mag}} ^{ij}$
using two trace-free combinations of the diagonal components,
$T^{\theta \theta}-T^{\phi \phi}$
and $2T^{rr}-T^{\theta \theta}-T^{\phi \phi}$.
These results are shown in the fourth and fifth panels of Figure~\ref{Fig.result1x5}. 
These components differ significantly in magnitude.
The magnetic components are approximately three orders of magnitude larger than the elastic components.
Therefore, a reduction factor of $10^{-3}$ is multiplied by 
 in the fourth and fifth panels, respectively.
The reason for this large difference is 
the hydrodynamic terms $T_{\rm{hyd}} ^{ij}$ and
$\delta \rho {\vec{\nabla}}\Phi_{\rm{G}}$ in Equation
(\ref{Forcebalance.eqn}).
Their ratio to the elastic term
is typically $p / \mu \sim 10^{3}$
 by the order-of-magnitude estimate.
The elastic stress does not necessarily compensate for the magnetic stress. 
The magnitude of the strain tensor $\sigma_{ij}$
is given by the sum of the axial and polar parts
\begin{equation}
  \frac{1}{2}\sigma_{ij} \sigma^{ij}
  = (\sigma_{\rm{ax}})^2+(\sigma_{\rm{po}})^2,
  \label{magofsigma}
\end{equation}
with their explicit forms written in terms of the combination of the components as
\begin{align} 
(\sigma_{\rm{ax}})^2 &= (\sigma^{\theta \phi})^2+(\sigma^{r \phi})^2
\nonumber
\\
&= \frac{1}{4\mu^2}
\left( (T_{\rm{elas}}^{\theta \phi})^2+ 
(T_{\rm{elas}}  ^{r \phi})^2
\right),
\\
(\sigma_{\rm{po}})^2 &= 
 \frac{1}{12}(2\sigma^{rr} -\sigma^{\theta\theta}-\sigma^{\phi\phi})^2
+\frac{1}{4}(\sigma^{\theta\theta}-\sigma^{\phi\phi})^2
+(\sigma^{r\theta})^2
\nonumber
\\
&= 
 \frac{1}{48\mu^2}(2T_{\rm{elas}}^{rr} -T_{\rm{elas}}^{\theta\theta}-T_{\rm{elas}} ^{\phi\phi})^2
+\frac{1}{16\mu^2}(T_{\rm{elas}}^{\theta\theta}-T_{\rm{elas}}^{\phi\phi})^2
+\frac{1}{4\mu^2}(T_{\rm{elas}}^{r\theta})^2.
\label{sygmagpol.eqn}
\end{align}
In the axial part $\sigma_{\rm{ax}}$, 
the major component in the magnitude 
is $|T_{\rm{elas}} ^{\theta \phi}|$, which
differs significantly from $|T_{\rm{mag}} ^{\theta \phi}|$,
although $|T_{\rm{elas}} ^{r\phi}| \approx |T_{\rm{mag}}^{r \phi}|$.
In the polar part $\sigma_{\rm{po}}$, 
the first and second terms in Equation~(\ref{sygmagpol.eqn})
differ from those of the magnetic part.
If the diagonal components
($T_{\rm{elas}}^{rr} $, $T_{\rm{elas}}^{\theta\theta}$, 
$T_{\rm{elas}} ^{\phi\phi}$) are replaced by
($T_{\rm{mag}}^{rr} $, $T_{\rm{mag}}^{\theta\theta}$, 
$T_{\rm{mag}} ^{\phi\phi}$)
or
(${\hat{T}}_{\rm{mag}}^{rr} $, ${\hat{T}}_{\rm{mag}}^{\theta\theta}$, 
${\hat{T}}_{\rm{mag}} ^{\phi\phi}$), the magnitude
$\sigma_{\rm{po}}$ increases significantly,
even though $|T_{\rm{elas}} ^{r\theta}| \approx |T_{\rm{mag}}^{r \theta}|$.
Thus, the magnitude of the 'elastic' tensor approximated by $T_{\rm{mag}} ^{ij}$ is significantly larger than that of the true elastic tensor.
%%

%%%%%%%%%%%%%%%%%[FIG3]%%%%%%%%%%%%%%%%%%%%%%%%
\begin{figure}\begin{center}
\includegraphics[width=\columnwidth]{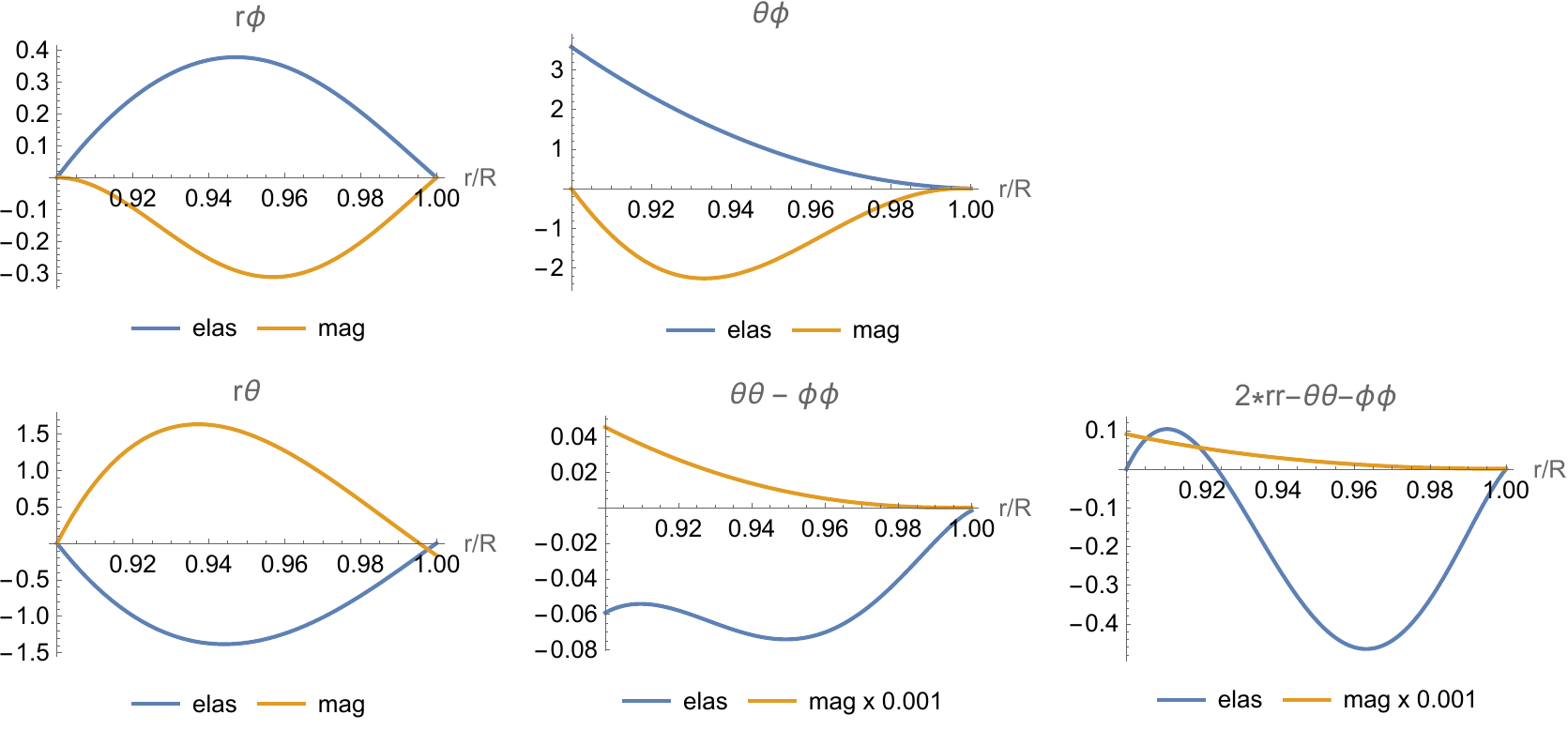}
\caption{ 
 \label{Fig.result2x5}
The radial functions for the elastic stress (in blue) and those for the magnetic stress
(in orange) are shown
for $T^{r \phi}$, $T^{\theta \phi}$, $T^{r \theta}$, 
$T^{\theta \theta}-T^{\phi \phi}$,
and $2T^{rr}-T^{\theta \theta}-T^{\phi \phi}$.
A factor of $10^{-3}$ is multiplied 
for 
$T_{\rm{mag}} ^{\theta \theta}-T_{\rm{mag}} ^{\phi \phi}$
and
$2T_{\rm{mag}} ^{rr}-T_{\rm{mag}} ^{\theta \theta}-T_{\rm{mag}} ^{\phi \phi}$. 
}
\end{center}\end{figure}
%%%%%%%%%%%%%%%%%%%%%%%%%%%%%%%%%%%%%%%%%%%%%%%

%%%%%%%%%%%%%%%%%[FIG4]%%%%%%%%%%%%%%%%%%%%%%%%
\begin{figure}\begin{center}
\includegraphics[width=\columnwidth]{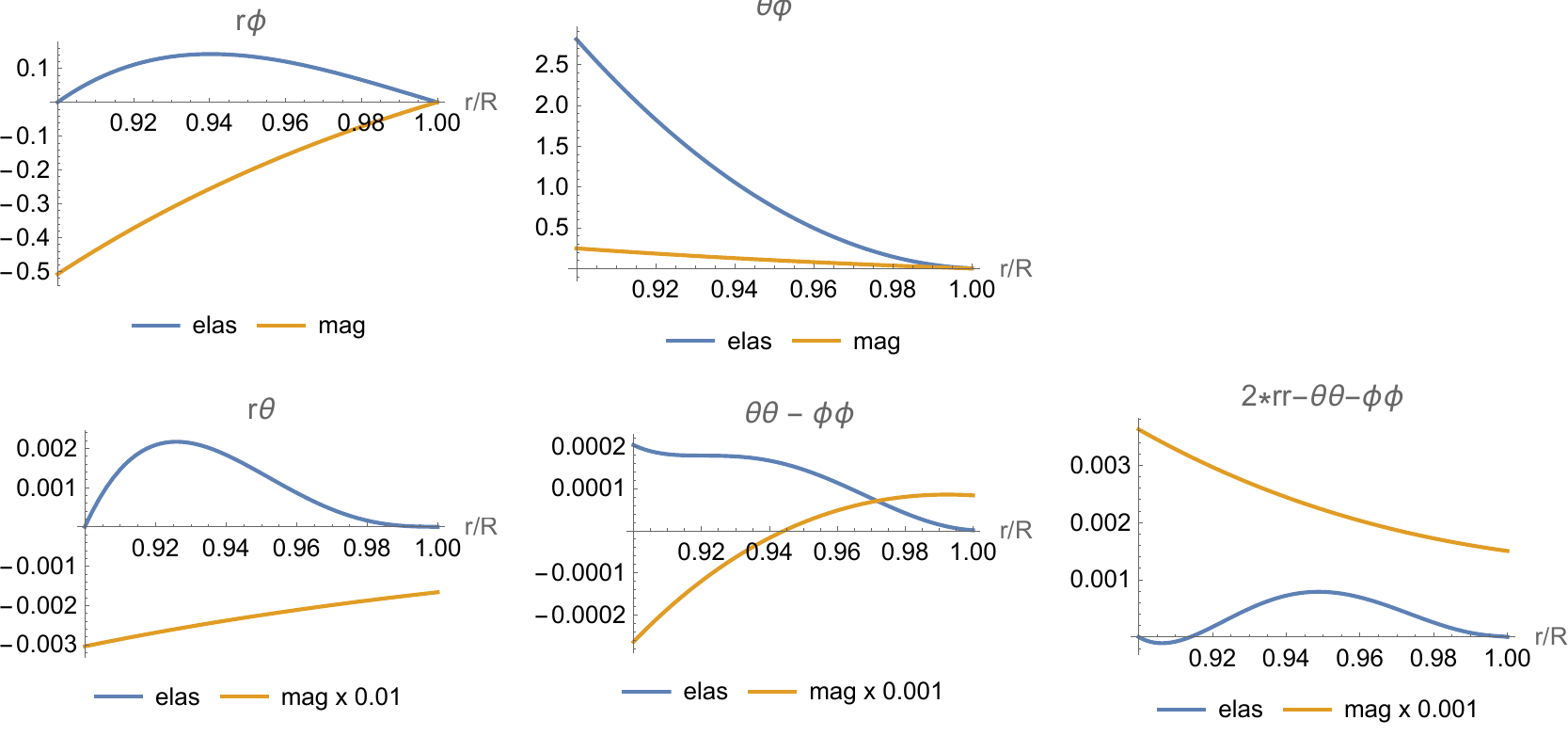}
\caption{ 
 \label{Fig.result3x5}
Comparison of the radial function for magnetic model II.
The radial function for the magnetic part (in orange)
in the bottom row is multiplied by 
a factor of $10^{-2}$, $10^{-3}$, and $10^{-3}$.
}
\end{center}\end{figure}
%%%%%%%%%%%%%%%%%%%%%%%%%%%%%%%%%%%%%%%%%%%%%%%

The angular dependence is definite for each tensor component,
$(T^{r \theta}, T^{r \phi})$ $\propto P_{l,\theta}$,
$(T^{\theta \phi}, T^{\theta \theta}-T^{\phi \phi})$ $\propto  W_{l}$,
and $2T^{rr}-T^{\theta \theta}-T^{\phi \phi}$ $\propto P_{l}$.
Apart from the trivial angular dependence, 
radial functions are more suitable for understanding 
the difference between elastic and magnetic stresses.
For the magnetic field Model I, 
the spherical harmonic index is given by $l=2$ only, except for
 $2T^{rr}-T^{\theta \theta}-T^{\phi \phi}$.
The latter is a mixture of $l=0$ and $l=2$; however, 
the monopole part of $T_{\rm{elas}} ^{ij}$
is considerably smaller than that of the quadrupole part, as shown by 
the angular dependence of $P_{2}$ 
provided in the fifth panel of Figure~\ref{Fig.result1x5}.
Figure~\ref{Fig.result2x5} shows their radial functions for $l=2$.
The approximation $T_{\rm{elas}} ^{ij} \approx -T_{\rm{mag}} ^{ij}$
is good for the $r\phi$ and $r\theta$ components,
 as shown in Figure~\ref{Fig.result1x5}. 
The comparison in the radial profile might be meaningless for the diagonal parts, i.e.,
$T^{\theta \theta}-T^{\phi \phi}$,
and $2T^{rr}-T^{\theta \theta}-T^{\phi \phi}$,
because their magnitudes are considerably different for the elastic and magnetic parts.
Figure~\ref{Fig.result3x5} shows the radial functions for Model II,
in which the magnetic field is continuous to the core $(r\le r_{c})$ and,
therefore, the values at $r_c$ change.
Unlike Figure~\ref{Fig.result2x5},
approximation $T_{\rm{elas}} ^{ij}  \approx -T_{\rm{mag}} ^{ij}$
does not hold for any of the components;
it is invalid for the $r \phi$ and $r\theta$ components.
This is owing to the boundary value at $r_c$;
 $T_{\rm{mag}} ^{ri} \ne 0~(i=\theta, \phi)$ in Model II,
whereas $T_{\rm{mag}} ^{ri}=0$ in Model I.
The elastic stress $T_{\rm{elas}} ^{ri}~(i=\theta, \phi)$ must disappear, 
 making the accidental coincidence $T_{\rm{mag}} ^{ri} =0~(i=\theta , \phi)$
 at $r_c$ necessary for the approximation.
$T_{\rm{elas}} ^{ij} \sim -T_{\rm{mag}} ^{ij}$
is useless as an estimate in the polar component because the magnitude is significantly different;
the elastic stress is two or three orders of magnitude smaller
than that of the magnetic stress, as shown by the three panels at the bottom row of Figure~\ref{Fig.result3x5}.
The elastic response to the magnetic force is minor in
the diagonal component, owing to the presence of hydrodynamic stress $T_{\rm{hyd}} ^{ij}$.
For the off-diagonal component $T_{\rm{elas}} ^{r\theta}$,
it is coupled with the diagonal component $T_{\rm{mag}} ^{ii}$
through the function $Y_l$ (see Equation~(\ref{tmn12elas.eqn})).
Consequently, a significant imbalance occurs between the elastic and magnetic stresses, and
the relationship $T_{\rm{elas}} ^{ij} \sim -T_{\rm{mag}} ^{ij}$ 
 is unreliable, even on the order of the estimate.
%%

%%%%%%%%%%%%%%%%%[FIG5]%%%%%%%%%%%%%%%%%%%%%%%%
\begin{figure}\begin{center}
\includegraphics[width=\columnwidth]{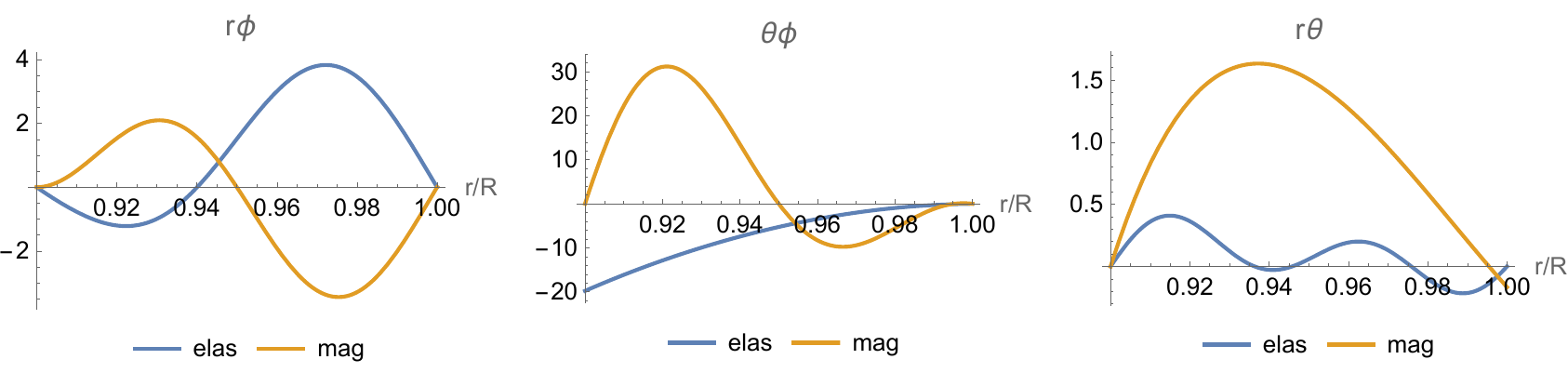}%
\caption{ 
\label{Fig.result5x3}
Radial functions for off-diagonal components are compared for Model III.
The elastic stress is shown in blue, and the magnetic stress is in orange.
}
\end{center}\end{figure}
%%%%%%%%%%%%%%%%%%%%%%%%%%%%%%%%%%%%%%%%%%%%%%%

Next, magnetic Model III is considered, in which the poloidal field is the same as that in Model I. 
However, the toroidal component is stronger and contains a radial node, as shown in the third panel of Figure ~\ref{Fig.Bgeometry}.
The poloidal and toroidal components are comparable in the magnitude.
The boundary values of $T_{\rm{mag}} ^{ij}$ both at $r=r_c$ and $R$ are the same as those for Model I.
Figure~\ref{Fig.result5x3} shows the
radial functions for the off-diagonal components.
In the $r\phi$ component, the approximation
  $T_{\rm{elas}}^{r\phi} \approx  -T_{\rm{mag}} ^{r\phi}$ holds, despite the radial node in $T_{\rm{mag}} ^{r\phi}$.
The approximation worsens in the $r\theta$-component.
The radial function for the magnetic part is 
the same as that for Model I, as shown in Figure~\ref{Fig.result2x5}
because $T_{\rm{mag}} ^{r\theta} \propto B^r B^\theta$ is unchanged from that of Model I.
However, 
the resultant $T_{\rm{elas}}^{r\theta}$ shown in Figure~\ref{Fig.result5x3}
 differs significantly from that in Figure~\ref{Fig.result2x5};
a wavy structure appears in the radial direction because of $B_{\phi}$ and
the absolute magnitude of the elastic function is reduced.
The polar part, $T_{\rm{elas}}^{r\theta}$,
is significantly affected by the significant strength of $B_{\phi} ^2$.
%

%%%%%%%%%%%%%%%%%[FIG6]%%%%%%%%%%%%%%%%%%%%%%%%
\begin{figure}\begin{center}
\includegraphics[width=0.7\columnwidth]{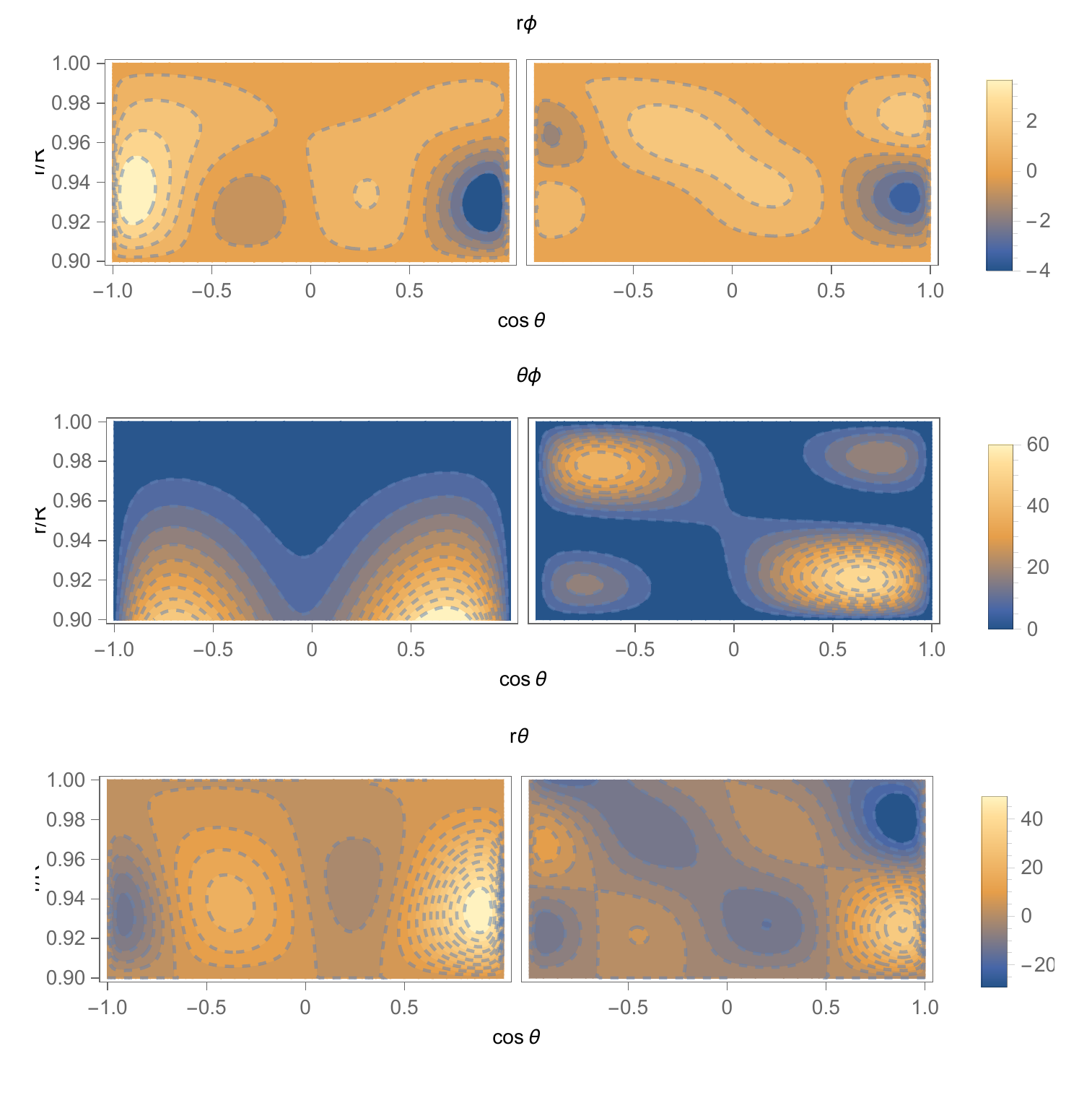}%
\caption{ 
\label{Fig.result6x3}
Display of stress components
by a contour map in the $\cos \theta$--$r/R$ plane.
The elastic stress $T_{\rm{elas}} ^{ij}$ (left panel)
is compared with the magnetic stress $ -T_{\rm{mag}} ^{ij}$ (right panel)
for Model IV.
From top to bottom,
$T ^{r\phi}, T^{\theta \phi}$, and $T^{r\theta}$ are shown.
}
\end{center}\end{figure}
%%%%%%%%%%%%%%%%%%%%%%%%%%%%%%%%%%%%%%%%%%%%%%%

\begin{figure}\begin{center}
\includegraphics[width=\columnwidth]{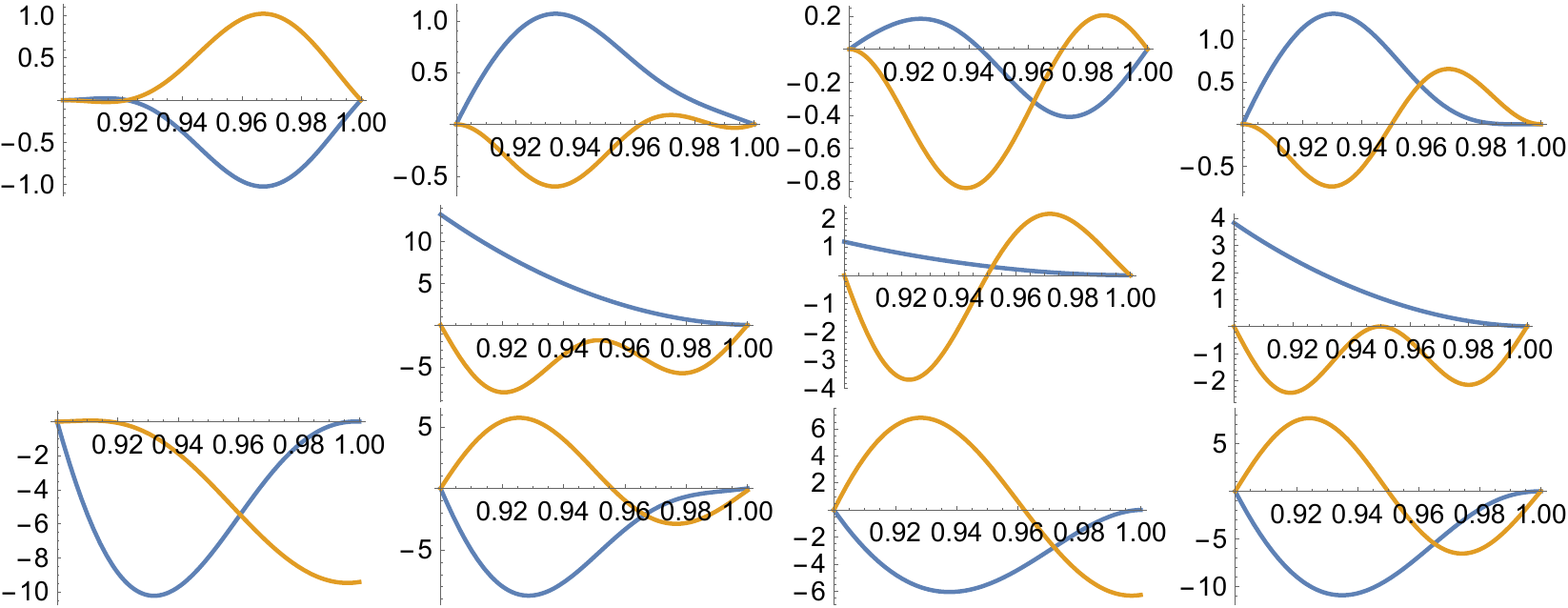}%
\caption{ 
 \label{Fig.result7x11}
The $r \phi$, $\theta \phi$, and $r \theta$ components for Model IV are shown from the top to bottom rows.
The spherical harmonics index from the left to right panels corresponds to $l=1$ to 4.
}
\end{center}\end{figure}

%%
A more complicated geometry of the magnetic field is considered, as
shown in the fourth panel of Figure~\ref{Fig.Bgeometry}(Model IV).
The poloidal field is a mixture of dipole $(l=1)$ and a confined component of $l=2$.
The toroidal field also comprises confined components with $l=$ 1 and 2.
The magnetic field is expelled from the core, and 
the boundary values for $B_r$ and $B_{\phi}$
 are the same as those for Model I.
The elastic and magnetic stresses in the off-diagonal components were compared.
Figure~\ref{Fig.result6x3} shows 
$T_{\rm{elas}} ^{ij} $ and $ -T_{\rm{mag}} ^{ij}$
 via contour maps in the $\cos \theta$--$r/R$ plane.
The maximum and minimum values of
$T_{\rm{elas}} ^{ij}$ approximately trace those of $-T_{\rm{mag}} ^{ij}$.
The radial profiles of the elastic stress are smooth.
For example, the two peaks of the magnetic stress in the radial direction are not clearly visible
in the elastic stress.
Therefore, examining the accuracy of
 $T_{\rm{elas}} ^{ij} \approx -T_{\rm{mag}} ^{ij}$ by eye is no longer useful.
 Figure~\ref{Fig.result7x11} shows radial functions
 decomposed using angular functions
 $(T^{r \theta}, T^{r \phi})$ 
$\propto P_{l,\theta}~(l=1, \cdots, 4)$,
$T^{\theta \phi}$ $\propto  W_{l}~(l=2, \cdots, 4)$.
In the $r\phi$ component (the first row of Figure~\ref{Fig.result7x11}),
all functions for the elastic and magnetic stresses 
vanish at both boundaries.
However, the elastic part does not respond locally to the 
magnetic part in the internal region, except for $l=1$.
Thus, their spatial profile as a sum of these functions
becomes significantly different.
In the $\theta \phi$ component (the second row of Figure~\ref{Fig.result7x11}),
the radial behavior is different in the elastic and magnetic parts;
the elastic functions decrease monotonically, whereas
the magnetic functions are sinusoidal.
This causes a significant difference in 
the second panel of Figure~\ref{Fig.result6x3}.
Finally, in the $r\theta$ component (the third row of Figure~\ref{Fig.result7x11}),
the boundary values at $r=R$ differ for $l=1$ and 3, and 
the elastic function is less oscillatory.
Thus, the approximation 
$T_{\rm{elas}} ^{ij} \approx -T_{\rm{mag}} ^{ij}$
is not justified or requires extremely restrictive conditions.

%%%%%%%%%%%%%%%%%[FIG7]%%%%%%%%%%%%%%%%%%%%%%%%
%%%%%%%%%%%%%%%%%%%%%%%%%%%%%%%%%%%%%%%%%%%%%%%

%3.2%%%%%%%%%%%%%%%%%%%%%%%%%%%%%%%%%%%%%%%%%%%
  \subsection{Magnitude of the strain tensor}
%%%%%%%%%%%%%%%%%%%%%%%%%%%%%%%%%%%%%%%%%%%%%%%
%%%%%%%%%%%%%%%%%[FIG8]%%%%%%%%%%%%%%%%%%%%%%%%
\begin{figure}\begin{center}
\includegraphics[width=0.6\columnwidth]{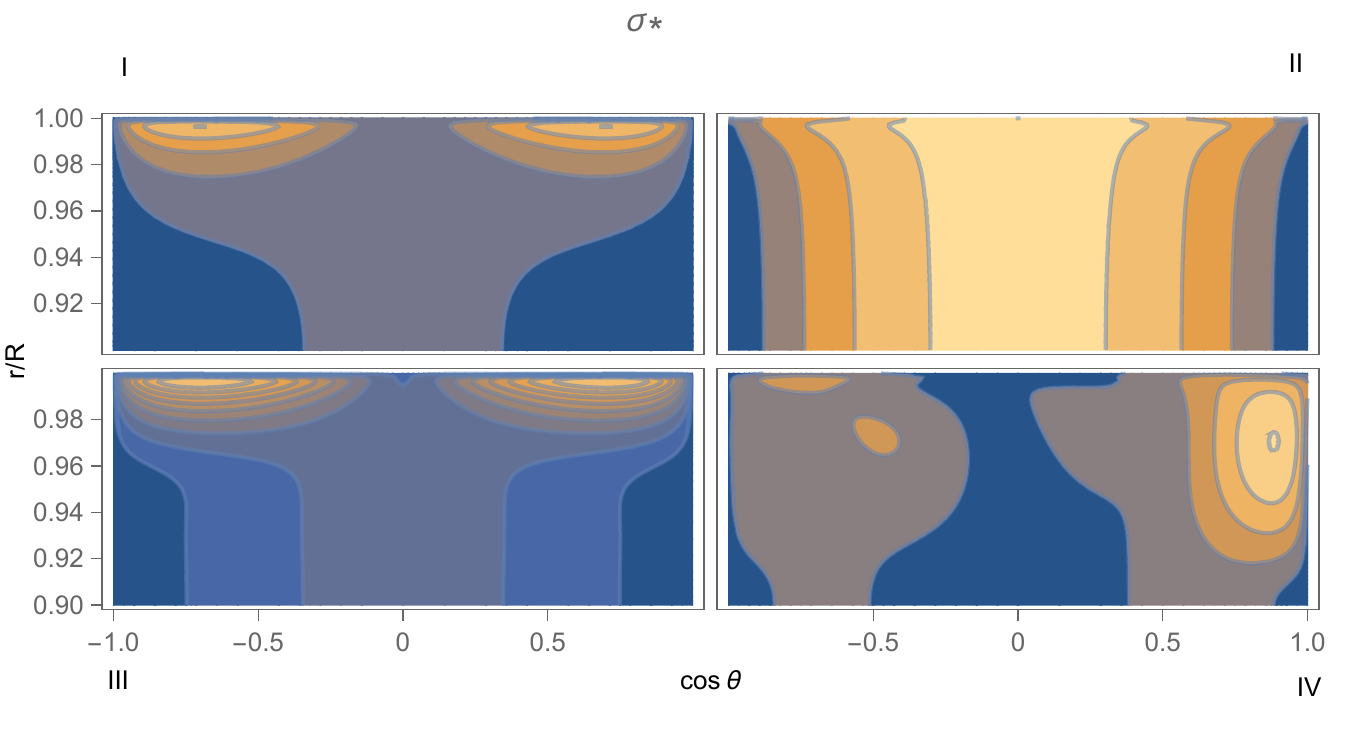}%
\caption{ 
 \label{Fig.sigma8}
The magnitudes of the strain tensor normalized by 
the maximum are shown 
on the $\cos \theta$ -- $r/R$ plain for Models I--IV.
}
\end{center}\end{figure}
%%%%%%%%%%%%%%%%%%%%%%%%%%%%%%%%%%%%%%%%%%%%%%%

It is useful to show the spatial distribution of the strain tensor magnitude $(\sigma _{ij} \sigma ^{ij}/2)^{1/2}$ (Equation~(\ref{magofsigma})),
 which is relevant to crustal failure according to the von Mises criterion.
The quantity indicates which part of the crust
 is broken under various magnetic fields.
Figure~\ref{Fig.sigma8} shows the normalized quantity
$\sigma_{*}$, for which the maximum value is unity.
The angular dependence in Model II (top right panel in Figure~\ref{Fig.sigma8}) 
originates from $W_{2}(\propto \sin^2 \theta)$ because the component
$\sigma^{\theta \phi} (\propto W_{2})$ is dominant.
The slightly complicated angular dependence in Models I and III 
are superpositions of $\sigma^{\theta \phi} \propto W_{2}$
and $(\sigma^{r \theta}, \sigma^{r \phi}) \propto \sin \theta \cos \theta$.
In Model IV, different $l$ modes are involved, and the spatial profile 
of $\sigma_{*}$ is more complex.
The maximum of $\sigma_{*}$ is located near the surface,
or the position is shifted to the outer part compared with that of $|T_{\rm{ela}} ^{ij}|$
because the shear modulus $\mu$ decreases with the radius.
The stable range of the crust
is limited as $(\sigma _{ij} \sigma ^{ij}/2)^{1/2} < \sigma_{c}$,
where $ \sigma_{c}$ is a number $\sigma_c \approx 10^{-2} -10^{-1}$
\citep{2009PhRvL.102s1102H,2018MNRAS.480.5511B,2018PhRvL.121m2701C}.
This condition results in the upper limit of the field strength
because $ \sigma _{ij} \propto B_{0}^2/ \mu_{c}$.
The overall magnetic field strength $B_{0}$
is constrained as
$ B_{0} < N_{a} \left( \sigma_{c}/0.1\right)^{1/2} \times 10^{14} {\rm{G}}$,
where $N_{a}$ is a number that depends on the models,
$N_{\rm{I}}=2.3$, $N_{\rm{II}}=5.1$, $N_{\rm{III}}=0.98$, and $N_{\rm{IV}}=1.1$.
Similar constraints for the overall magnetic field strength
are also obtained using different 
numerical methods for various magnetic fields~\citep{2022MNRAS.511..480K,2023MNRAS.519.3776F}.
The precise value $N_{a}$ and the failure position
are determined by solving the differential equations for the
magnetic field geometry, although the magnitude 
$N_{a}={\mathcal{O}}(1)$ is simply estimated.
%%

%(4)%%%%%%%%%%%%%%%%%%%%%%%%%%%%%%%%%%%%%%%%%%%
\section{Discussion}
%%%%%%%%%%%%%%%%%%%%%%%%%%%%%%%%%%%%%%%%%%%%%%%
The elastic deformation of the crust due to magnetic stress is considered.
The crustal strain was obtained by solving the differential equations 
with the appropriate boundary conditions. 
This study aims to compare the results with those obtained using the algebraic approximation in previous studies without any justification.
The validity of the algebraic approximation was tested.
The results of this study demonstrate that the algebraic expression never represents the correct elastic tensor.
The resultant profile of the elastic stress is smoother than the magnetic one given as the source term.
The radial variation is remarkable owing to the thin crust; however, it becomes less sharp
 when the differential equations are solved;
 the source term spreads globally in the solution.
The magnitude of the elastic stress is generally smaller than that of the magnetic stress.
 The elastic stress is significantly smaller, especially in the diagonal components.
Therefore, the elastic stress tensor cannot be approximated by
any algebraic expression from the magnetic stress.
Moreover, it is necessary to revisit 
evolution calculations which used the spurious criteria~\citep{
2011ApJ...727L..51P,2013MNRAS.434..123V,2019MNRAS.486.4130L,
2020ApJ...902L..32D,2021MNRAS.506.3578G}.
The present method for calculating the elastic tensor
will be useful for improving the models.
In numerical simulation of the magnetic field evolution, 
$\sigma ^{ij}$ is monitored by solving differential equations.
Compared with an algebraic relation, this method significantly increases computing time, 
but is an inevitable step for examining the elastic limit of the crust.
%%

%%%
The reason the algebraic expression disagrees with
the correct result is discussed here.
In the axially symmetric model,
two components $\sigma^{r\phi}$ and $\sigma^{\theta \phi}$
relevant to the axial part
are described by a single function $\xi_{\phi}$, and
satisfy
$\sin \theta (\sigma^{r\phi}/\sin \theta)_{, \theta}$
$=r(\sigma^{\theta \phi})_{,r}$
(see Equation~(\ref{tmnaxelas.eqn})).
The magnetic field is conveniently expressed by two functions
$\Psi$ and $S(\equiv r\sin \theta B_{\phi})$, with the
two components of the magnetic stress given by
$(T_{\rm{mag}}^{r\phi}, T_{\rm{mag}}^{\theta\phi})$
$=S/(4\pi r^2 \sin^2\theta)[{\VEC{\nabla}} \Psi \times \VEC{e}_{\phi}]$.
The differential relation in which 
$(\sigma^{r\phi}, \sigma^{\theta \phi})$ are substituted by 
$-(2\mu)^{-1}(T_{\rm{mag}}^{r\phi}, T_{\rm{mag}}^{\theta\phi})$
does not holds, or strongly constrains $\Psi$ and $S$~\footnote{
%%%
$T_{\rm{mag}}^{ij}$ is the difference between two states, and
the constraint is for the initial states ($\Psi$, $S$) 
and their changes ($\delta \Psi$, $\delta S$).
}.
Therefore,
the algebraic relation 
$T_{\rm{elas}}^{i\phi}+T_{\rm{mag}}^{i\phi}=0$
$(i=r, \theta)$ is not compatible.
The argument for the polar part is complex, but
the relation 
$T_{\rm{elas}}^{ij}+T_{\rm{mag}}^{ij}=0$
is satisfied only under restrictive conditions. 
One reason for this is that the material perturbation
$T_{\rm{hyd}}^{ij}$ is induced. 
The pressure is significantly greater than the elastic and magnetic forces, so
even a slight change in pressure perturbation significantly modifies the other forces.
When restricting the class of displacement to neglect material perturbation, 
the four components $\sigma^{rr}$, $\sigma^{\theta \theta}$, $\sigma^{\phi\phi}$,
and $\sigma^{r\theta}$ are expressed by two functions, 
$\xi_{r}$ and $\xi_{\theta}$, and are related to each other.
The magnetic tensors
$T_{\rm{mag}}^{rr}$, $T_{\rm{mag}}^{\theta\theta}$, $T_{\rm{mag}}^{\phi\phi}$,
and $T_{\rm{mag}}^{r\theta}$ are expressed by the 'square' of 
$\Psi$ and $S$.
Therefore, the linear relation 
$T_{\rm{elas}}^{ij}+T_{\rm{mag}}^{ij}=0$ is not globally satisfied.

The elastic deviation due to magnetic stress is widespread, and is described by a solution of differential equations to be fitted to boundary conditions. The elastic response is not approximated by solely considering the local magnetic stress,
however the explicit calculations given in this study are inevitable. 
This study provided the correct crustal failure process according to the evolution of the magnetic field.

%%%%%%%%%%%%%%%%%%%%%%%%%%%%%%%%%%%%%%%%%%%%%%%%%%
\section*{Acknowledgments}
%%%%%%%%%%%%%%%%%%%%%%%%%%%%%%%%%%%%%%%%%%%%%%%%%%
 This work was supported by JSPS KAKENHI Grant Numbers JP19K03850, JP23K03389.
%%%%%%%%%%%%%%%%%%%%%%%%%%%%%%%%%%%%%%%%%%%%%%%%%%
%%%%%%%%%%%%%%%%%%%% REFERENCES %%%%%%%%%%%%%%%%%%
%% for ApJ
%% below two
 \bibliography{kojima24Aug}
\bibliographystyle{aasjournal} 
%%%%%%%%%%%%%%%%%%%%%%%%%%%%%%%%%%%%%%%%%%%%%%%%%%
%%%%%%%%%%%%%%%%%%%%%%%%%%%%%%%%%%%%%%%%%%%%%%%%%%
    \end{document}